\documentclass[
    ,final            
    ,numberedheadings 
  ]
  {aipproc}

\layoutstyle{8x11single}
\usepackage{graphicx,color}
\usepackage[hang]{subfigure}
\usepackage[intlimits,fleqn]{amsmath}
\usepackage{amsfonts}
\usepackage{amsbsy}
\usepackage{amssymb,dcolumn,exscale}
\usepackage{cite}
\graphicspath{{pics}}

\newcommand{\Cf}{C_F}
\newcommand{\Ca}{C_A}
\newcommand{\nf}{n_f}
\newcommand{\as}{\alpha_s}

\newcommand{\half}{\ensuremath{\tfrac{1}{2}}}
\newcommand{\mq}{m_{\tilde{q}}}

\newcommand{\mg}{m_{\tilde{g}}}
\newcommand{\sq}{\tilde{q}}
\newcommand{\sqb}{\tilde{q}^{\ast}}

\newcommand{\dif}{\ensuremath{\mathrm{d}}}

\newcommand{\GeV}{\ensuremath{\,\mathrm{GeV}}}
\newcommand{\TeV}{\ensuremath{\,\mathrm{TeV}}}
\newcommand{\fb}{\ensuremath{\,\mathrm{fb}}}
\newcommand{\pb}{\ensuremath{\,\mathrm{pb}}}

\newcommand{\barq}{\ensuremath{\bar{q}}}
\newcommand{\hats}{\hat{s}}
\newcommand{\agg}{a_1^{gg}}
\newcommand{\aqq}{a_1^{qq}}
\def\shat{{\hat s}}
\def\muf{{\mu^{}_f}}

\def\mur{{\mu^{}_r}}

\DeclareMathOperator*{\Li}{Li_2}

\newcommand{\ent}{\ensuremath{\stackrel{\wedge}{=}}}

\allowdisplaybreaks[2]
\renewcommand{\tablename}{FIGURE} 

\begin{document}

\title{
\textnormal{\normalsize
            \phantom{m}\\[-15mm]
            DESY-09-122\phantom{MMMMMMMMMMMMMMMMMMMMMMMMMMMMMMMMMMMMMMMMMMM}\\[0mm]
            SFB/CPP-09-70\phantom{MMMMMMMMMMMMMMMMMMMMMMMMMMMMMMMMMMMMMMMMMM}\\[2mm]
            }
    Squark pair production at the LHC}


\author{U. Langenfeld}{
  address={Deutsches Elektronensynchrotron DESY, Platanenallee 6, D--15738 Zeuthen}
}

\begin{abstract}
We present NNLO cross sections for squark- antisquark production at the LHC.
We have calculated new analytic expressions for the scale dependent scaling
functions at one and two loop.
\end{abstract}

\maketitle


\section{Squark pair production cross section at the LHC}
If Supersymmetry is realised in Nature then it is expected that squark and gluino 
pairs are produced in large numbers at the LHC.
It is possible to probe masses up to the $\TeV$ range. 
Squarks are assumed to be heavier than $\approx 400\GeV$~\citep{:2007ww} 
so these particles are produced near the kinematical production threshold. 
Therefore one can use the same methods to calculate higher 
order cross sections as developed for $t\bar{t}$ 
production~\citep{Kidonakis:2001nj,Moch:2008qy}.
The partonic LO and NLO cross sections are known for long times~\citep{Beenakker:1996ch}.
Approximate NNLO corrections have been calculated in~\citep{Langenfeld:2009eg}.
The LO partonic cross section and  the NLO theshold expansion are known
analytically~\citep{Beenakker:1996ch}.
In this article we present analytical formulae for the scale dependence determining 
NLO scaling functions and the threshold expansion of the scale dependence determining 
NNLO scaling functions. 
For related work, see also Ref.~\citep{Kulesza:2008jb}.
In Ref.~\citep{Beneke:2009rj}, the soft anomalous dimension has been calculated to 
NNLO accuracy.

The partonic cross section $\shat$ with identified renormalisation and 
factorisation scale can be expanded as 
\begin{equation}
 \hat{\sigma}_{ij} = \frac{\as^2}{\mq^2}\biggl[f^{(00)}_{ij} 
                + 4\pi\as \Big(f^{(10)}_{ij} + f^{(11)}_{ij} L_M\Big) 
                + (4\pi\as)^2 \Big(f^{(20)}_{ij} + f^{(21)}_{ij} L_M 
                            + f^{(22)}_{ij} L_M^2\Big)\biggr],
                            \enspace L_M = \log(\mu^2/\mq^2),
\end{equation}
where $ij$ denote the initial states gluon - gluon or quark - antiquark.
The full dependence on the renormalisation and factorisation scale is the same as for 
$t\bar{t}$-production, see Ref.~\citep{Langenfeld:2009wd}.
The hadronic cross section is given as a convolution of the partonic cross section
with the parton luminosities $L_{ij}$:
\begin{equation}
 \sigma_{pp \to \sq\sqb X}(s,\mq,\mg) = \sum_{i,j = q,\barq,g}\quad\int_{4\mq^2}^{s}\!\!
                                        \dif\hats \,L_{ij} \,
                                        \hat{\sigma}_{ij}(\hats,\mq,\mg,\mu).
\end{equation}

We performed a scan of the LO, NLO, and NNLO squark pair production cross 
section in the $\mg$ - $\mq$ - plane using \texttt{Prospino}~\citep{Beenakker:1996ed} 
and the formulae presented in~\citep{Langenfeld:2009eg}, see 
Fig.~\ref{fig:tot}(a) -(c).
One clearly sees the strong enhancement of the NLO and NNLO cross section 
compared to the LO cross section: 
For squarks and a gluino with mass $200\GeV$ and $250\GeV$, respectively,
we have a LO cross section of about $500\pb$, but about $1000\pb$ at NLO and NNLO.
At NNLO, the $1000\pb$ region is even enlarged to higher gluino masses.
For a squark mass of $400\GeV$ and a gluino mass of $500\GeV$, we find
for the LO, NLO, and NNLO cross section $28.9\pb$, $43.1\pb$, and $46.9\pb$,
respectively. 
The NNLO cross section is $9\%$ larger than the NLO cross section.
The contour lines of constant cross section are running nearly parallel
to the gluino mass axis: 
The cross section shows a rather mild dependence on the gluino mass.
There is a weak enhancement of the cross section for $\mq = \mg$ as one can see
from the small bump at $\mg = \mq$.
The cross section decreases for more than three orders of magnitude
for squark masses from $200 - 1000\GeV$.
This strong mass dependence is well-known from hadronic $t \bar{t}$ pair production.

In Fig.~\ref{fig:tot}(d) we show the full $\muf$ - $\mur$ scale dependence of the NNLO
cross section for the example point $\mq = 400\GeV$, $\mg = 500\GeV$.
The scale uncertainty is about $-8\%$ for $(\muf,\mur) =(\half \mq, 2\mq)$ and
about $+8\%$ for $(\muf,\mur) =(2\mq, \half\mq)$.
This is considerably larger than the usual scale uncertainty taken at 
$\mur = \muf \equiv \mu$ (in our example $\approx -4\%$ at $\mu = 1/2$ and 
$+1\%$ at $\mu = 2$).
This shows that a full treatment of the scale dependence leads to more reliable
estimates of the scale uncertainty.

\begin{table}
\begin{tabular}{cc}
 \centering
    {
        \includegraphics[width=7.0cm]{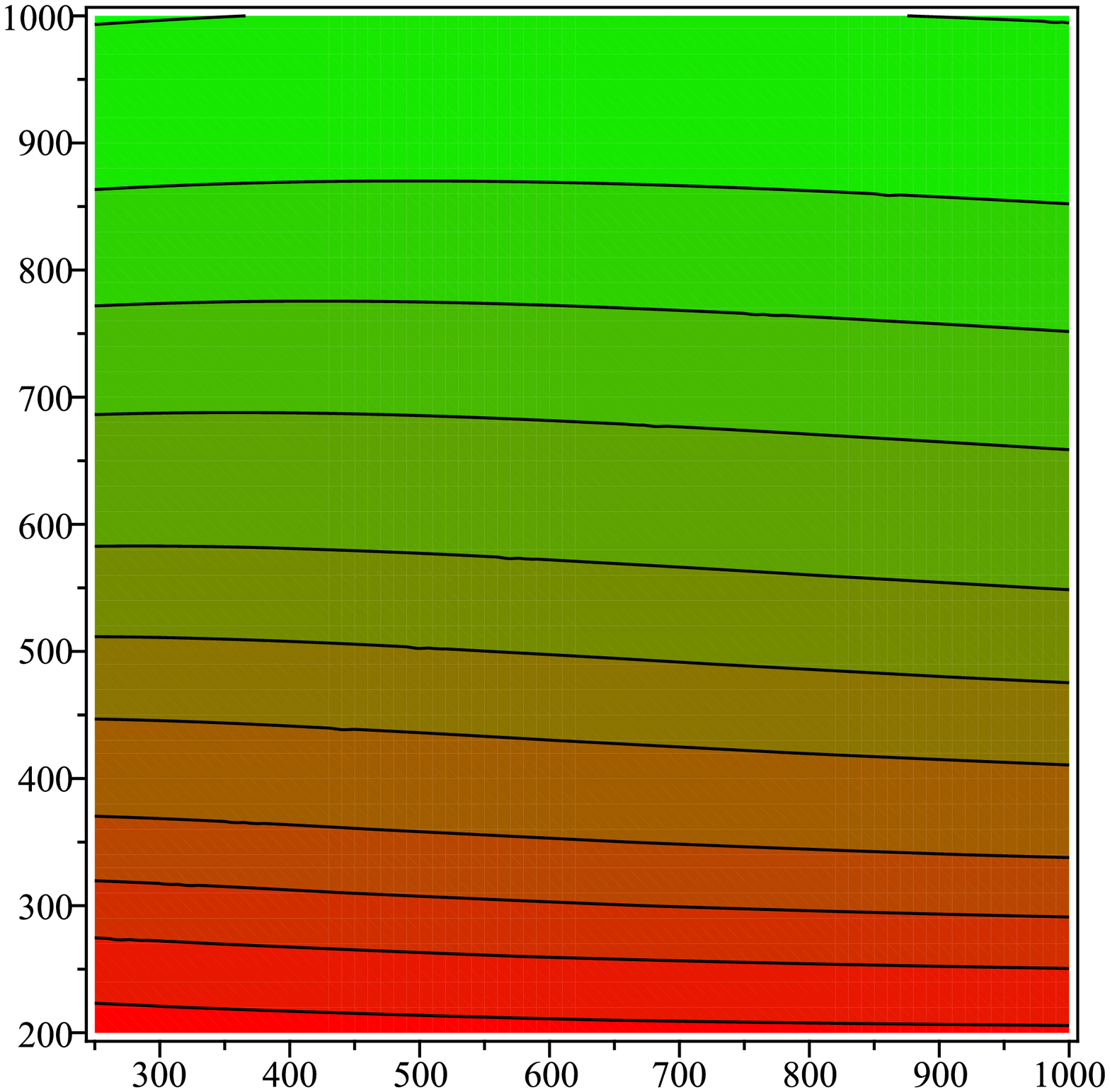}
        \put(-120,-6){\footnotesize{$\mg [\!\GeV]$}}
        \put(-204,95){\footnotesize{\rotatebox{90}{$\mq [\!\GeV]$}}}
        \put(-174,19){\footnotesize{$500\pb$}}
        \put(-174,31){\footnotesize{$200\pb$}}
        \put(-174,41){\footnotesize{$100\pb$}}
        \put(-174,52){\footnotesize{$\phantom{0}50\fb$}}
        \put(-174,69){\footnotesize{$\phantom{0}20\fb$}}
        \put(-174,85){\footnotesize{$\phantom{0}10\pb$}}
        \put(-174,101){\footnotesize{$\phantom{00}2\pb$}}
        \put(-174,125){\footnotesize{$\phantom{00}5\pb$}}
        \put(-174,144){\footnotesize{$\phantom{00}1\pb$}}
        \put(-174,164){\footnotesize{$\phantom{0}0.5\pb$}}
        \put(-174,185){\footnotesize{$\phantom{0}0.2\pb$}}
    }
    &
    {
        \includegraphics[width=7.0cm]{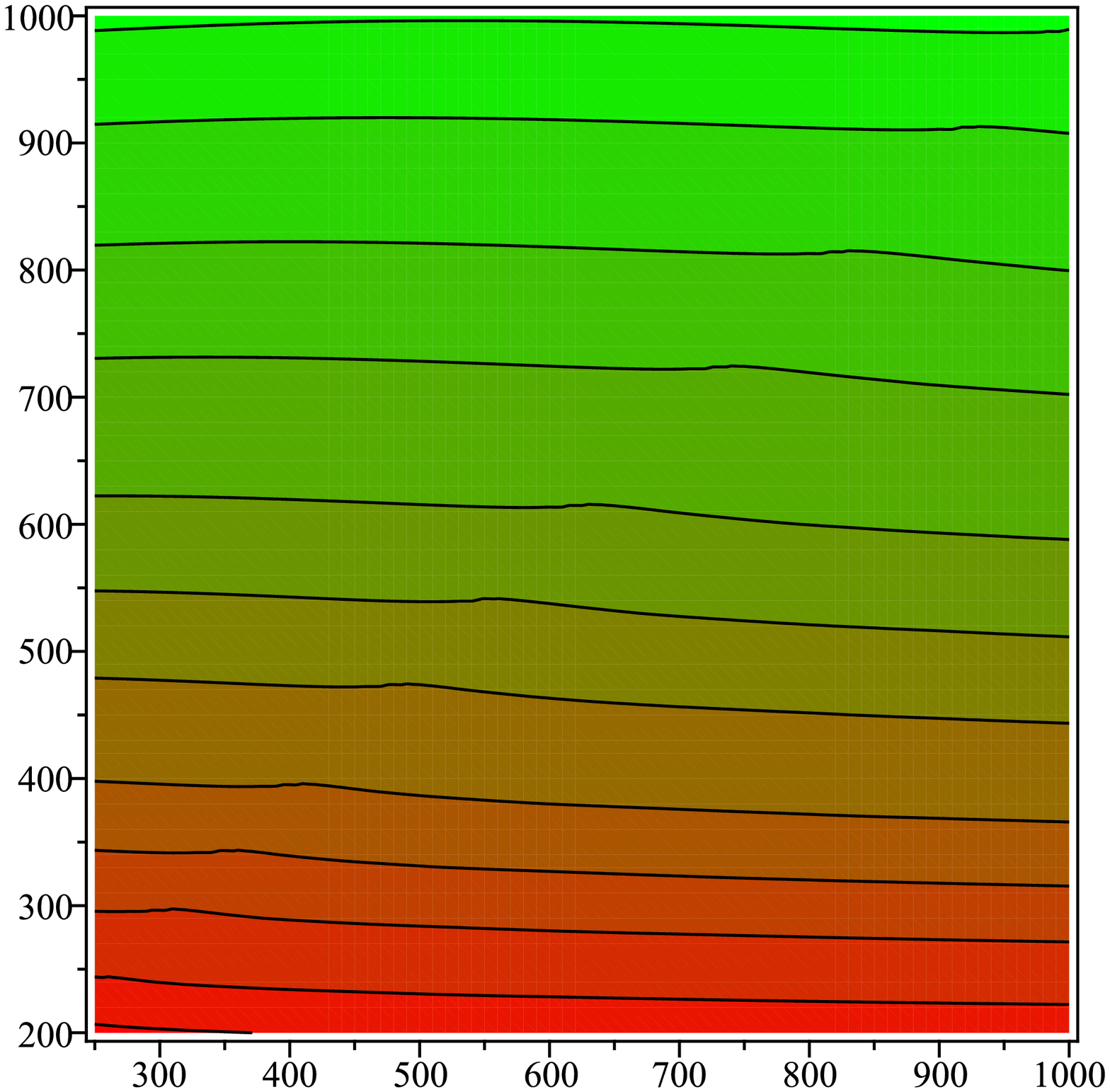}
        \put(-120,-6){\footnotesize{$\mg [\!\GeV]$}}
        \put(-204,95){\footnotesize{\rotatebox{90}{$\mq [\!\GeV]$}}}
        \put(-174,16){\footnotesize{$1000\pb$}}
        \put(-174,24){\footnotesize{$500\pb$}}
        \put(-174,37){\footnotesize{$200\pb$}}
        \put(-174,46){\footnotesize{$100\pb$}}
        \put(-174,59){\footnotesize{$\phantom{0}50\pb$}}
        \put(-174,77){\footnotesize{$\phantom{0}20\pb$}}
        \put(-174,93){\footnotesize{$\phantom{0}10\pb$}}
        \put(-174,109){\footnotesize{$\phantom{00}5\pb$}}
        \put(-174,133){\footnotesize{$\phantom{00}2\pb$}}
        \put(-174,154){\footnotesize{$\phantom{00}1\pb$}}
        \put(-174,168){\footnotesize{$\phantom{0}0.5\pb$}}
        \put(-174,185){\footnotesize{$\phantom{0}0.3\pb$}}
    }
    \\[3mm]
    (a) & (b) \\[2mm]
    {
        \includegraphics[width=7.0cm]{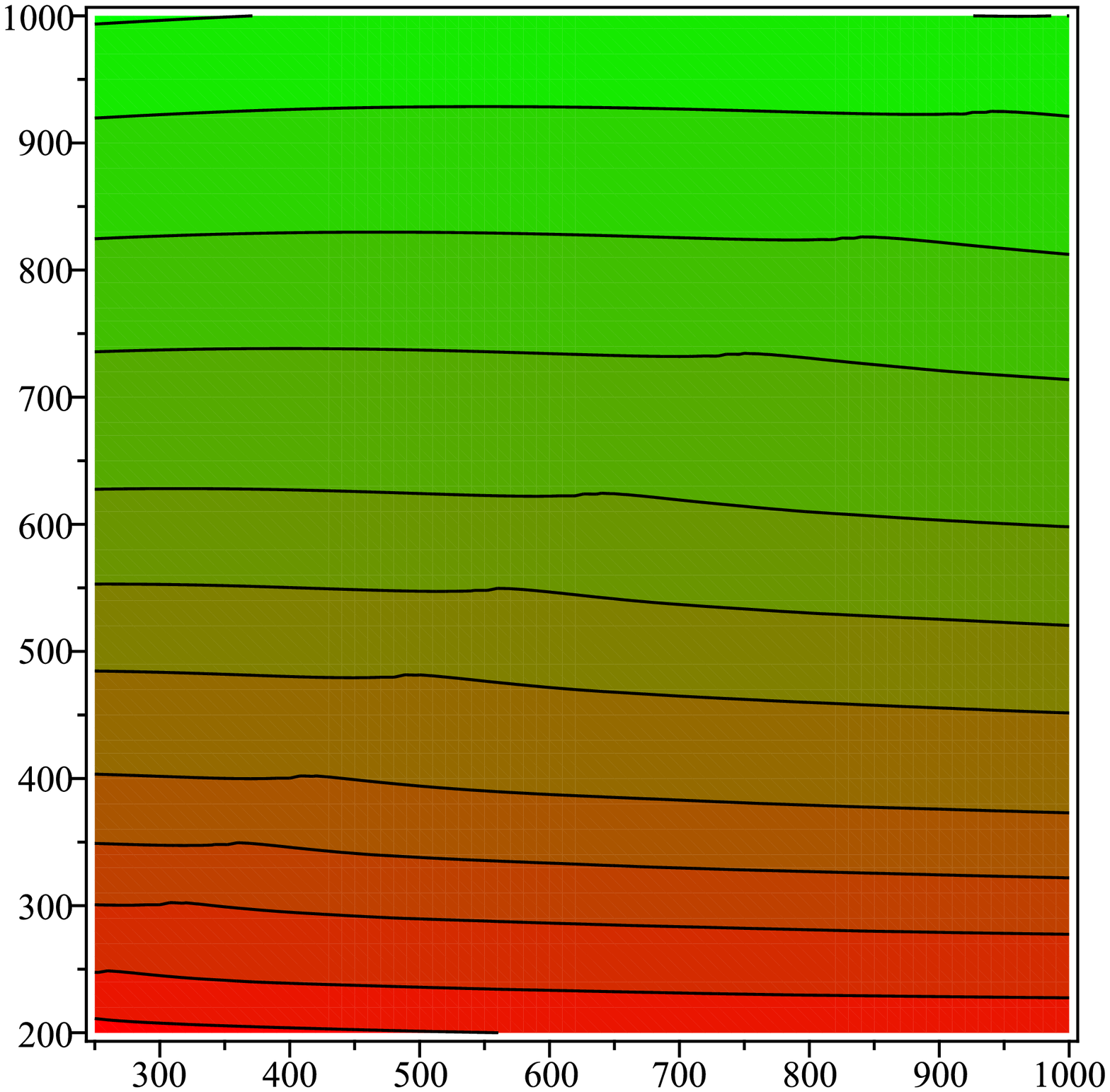}
        \put(-120,-6){\footnotesize{$\mg [\!\GeV]$}}
        \put(-204,95){\footnotesize{\rotatebox{90}{$\mq [\!\GeV]$}}}
        \put(-174,17){\footnotesize{$1000\pb$}}
        \put(-174,26){\footnotesize{$500\pb$}}
        \put(-174,37){\footnotesize{$200\pb$}}
        \put(-174,48){\footnotesize{$100\pb$}}
        \put(-174,61){\footnotesize{$\phantom{0}50\pb$}}
        \put(-174,78){\footnotesize{$\phantom{0}20\pb$}}
        \put(-174,94){\footnotesize{$\phantom{0}10\pb$}}
        \put(-174,111){\footnotesize{$\phantom{00}5\pb$}}
        \put(-174,135){\footnotesize{$\phantom{00}2\pb$}}
        \put(-174,155){\footnotesize{$\phantom{00}1\pb$}}
        \put(-174,170){\footnotesize{$\phantom{0}0.5\pb$}}
        \put(-174,185){\footnotesize{$\phantom{0}0.3\pb$}}
    }
    &
    {
        \includegraphics[width=7.0cm]{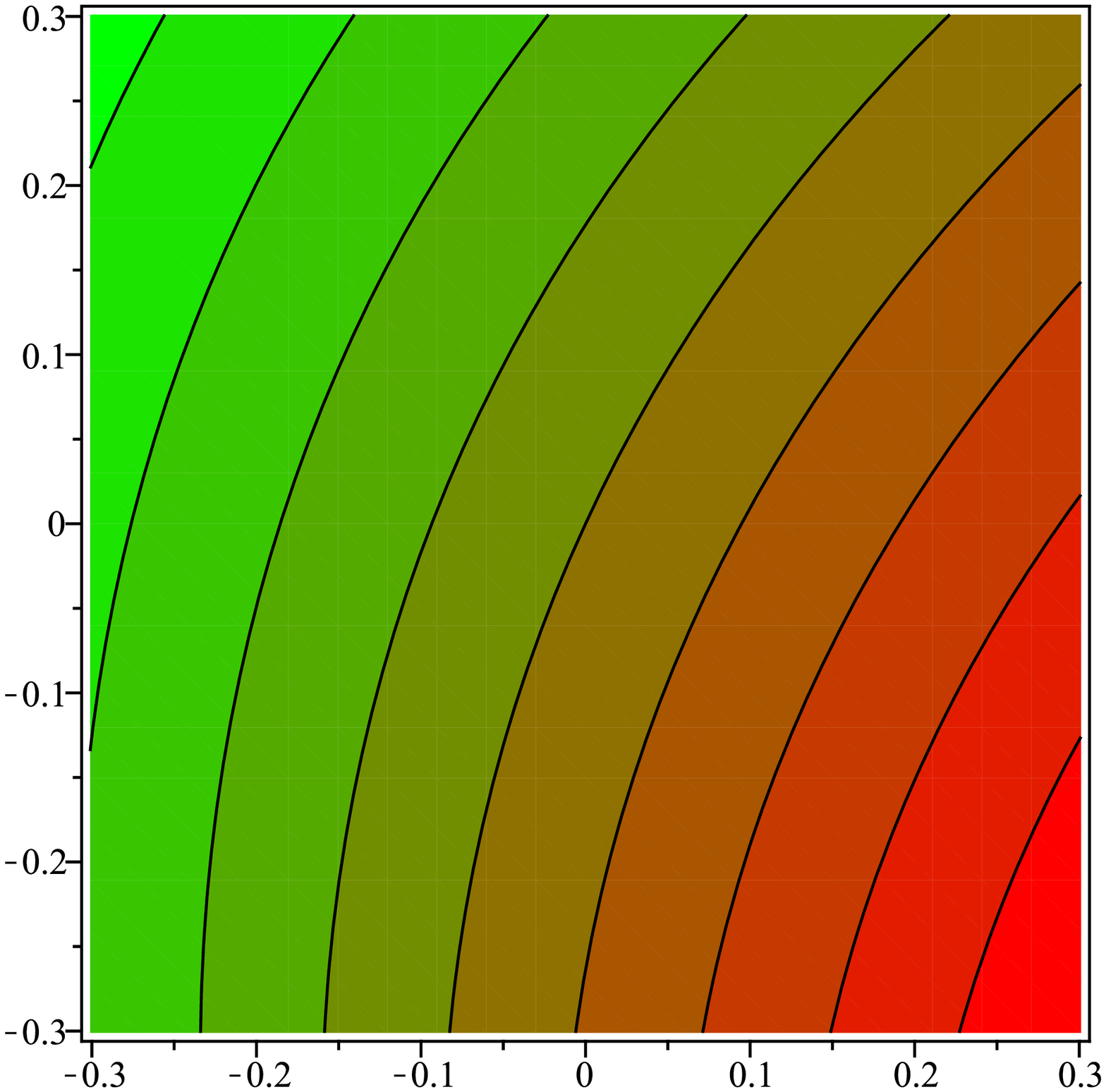}
        \put(-120,-6){\footnotesize{$\log_{10}(\mu_f)$}}
        \put(-204,95){\footnotesize{\rotatebox{90}{$\log_{10}(\mu_r)$}}}
        \put(-177,126){\rotatebox{68}{\footnotesize{$44.1\pb \ent -6\%$}}}
        \put(-153,120){\rotatebox{64}{\footnotesize{$45\pb \ent -4\%$}}}
        \put(-129,110){\rotatebox{64}{\footnotesize{$46\pb \ent -2\%$}}}
        \put(-102,100){\rotatebox{62}{\footnotesize{$46.9\pb$}}}
        \put(-87, 80){\rotatebox{64}{\footnotesize{$47.9\pb \ent + 2\%$}}}
        \put(-70, 60){\rotatebox{64}{\footnotesize{$48.8\pb \ent + 4\%$}}}
        \put(-54, 40){\rotatebox{64}{\footnotesize{$49.7\pb \ent + 6\%$}}}
        \put(-38, 15){\rotatebox{70}{\footnotesize{$50.7\pb \ent + 8\%$}}}
    }
    \\[3mm]
    (c) & (d)
\caption{\small LO (Fig.~(a)), NLO (Fig.~(b)), and NNLO (Fig.~(c))
        squark pair production cross sections at the LHC @ $14\,\TeV$.
        Figure (d) shows the $\mu_f$ - $\mu_r$ dependence of the NNLO
        cross section for $\mq = 400\GeV$ and $\mg = 500\GeV$.
        The PDF set is CTEQ6.6.}
 \label{fig:tot}
\end{tabular}
\end{table}

\section{Analytical Formulae}
In this section, we present analytical formulae for the $f^{(11)}_{ij}$ scaling functions
and the threshold logarithms of the $f^{(21)}_{ij}$ and $f^{(22)}_{ij}$ scaling functions.
The $f^{(11)}_{ij}$ scaling functions are determined by the renormalisation group equation:
\begin{equation}
\label{eq:f11}
 f^{(11)}_{ij} = \frac{1}{8\pi^2}\left(\beta_0 f^{(00)}_{ij} 
            - P_{ij}^{(0)}\otimes f^{(00)}_{ij}\right),
 \quad ij = gg,\enspace q\bar{q}.
\end{equation}
The $ P_{ij}^{(0)}$ are the leading order splitting functions, see Ref.~\citep{Moch:1999eb}.
$\otimes$ denotes the standard Mellin convolution.
The scale dependent NLO scaling function Eq.~\eqref{eq:f11} depends only on LO functions.
Performing the integrations yields as new analytic results the 
Eq.~\eqref{eq:fgg11} and \eqref{eq:fqq11}.

\begin{align}
\label{eq:fgg11}
f_{gg}^{(11)} \,=\, &-\frac{1}{384\*\pi}\*\Ca\*\nf\*\rho\*
                \Bigl[\beta \bigl(10 + 31\*\rho\bigr) L_3
               -\rho\*\bigl(\rho + 16 \*\bigr)\*\bigl(L_3\* L_2 + L_5\bigr) 
               +\rho\*\bigl(\rho - 16 \*\bigr)
               \bigl(L_4 - \tfrac{1}{2}\* L_6\bigr)\nonumber\\[2mm]
              &\hspace*{10mm}+\bigl(-20 +34\*\rho + \tfrac{127}{12}\*\rho^2 - 16\*\rho\*\log(2)
                + \rho^2\*\log(2)\bigr)L_2
               +\tfrac{1}{90}\*\beta\*\bigl(2606 - 14763\rho+ 352\*\rho^{-1}\bigr)
               \Bigr],\\[4mm]
\label{eq:fqq11}
f_{qq}^{(11)} \,=\, &\frac{1}{8\*\pi^2}\beta_0 f_{qq}^{(00)}
 -\frac{1}{216\*\pi}\*\Cf\*\nf\*\delta_{ij}\*\rho
    \Bigl[(3\*\rho -2)\*L_2 + 4\*\beta^3\*L_3 - \tfrac{1}{3}\*\beta\*(13 - 7\*\rho)\Bigr]
 \nonumber\\[2mm]
 &-\frac{1}{216\*\pi}\*\Cf\*\delta_{ij}\*\rho \*\biggl\{
 \rho\*(1+a)\*\Bigl[L_1^2 - \half \*L_1 + 2 L_1 \* L_2 
    + 2 L_1 \* L_3 + L_4 -\half L_6 - L_7 - L_8 -\log\left(\tfrac{1+a}{2}\right)L_2
     \nonumber\\[2mm]
              &\hspace*{75mm}
    - 4 \*\Li\left(-\tfrac{2\*\beta}{1-\beta}\right)
    + 4 \*\Li\left(-\tfrac{2 \*a \*\beta}{a\*(1-\beta)+2}\right)
    \Bigr]\nonumber\\[2mm]
   & \hspace*{8mm}
   +\half\rho^2\*a^2\*\Bigl[L_9 - L_5 + L_1 \* L_3\Bigr]
   +2\*(a \rho + 2)\*\Bigl[\beta\*L_3 - L_2\Bigr]
   -\frac{2}{a^2}\*(1 + 2\* a)\*L_2
   -\frac{2}{a^2}\*(1 + a)^2\*L_1
   +\frac{2}{a}\*\beta\*(1 - a)
      \biggr\}\nonumber\\[2mm]
 &+\frac{1}{72\*\pi}\* \rho\* \Cf\*\biggl\{
   -\bigl(a\* \rho + 2 \bigr)\*
        \Bigl[-2\*L_9 -L_4 +L_7 +L_8 + \tfrac{1}{2} \* L_6
               -2 L_1 \* L_3 +\log\left(\tfrac{1+a}{2}\right) \*L_2 
               + 4 \Li\left( - \tfrac{2\*\beta}{1-\beta} \right)\Bigr]\nonumber\\[2mm]
   &+\frac{1}{2\*(a+1)\*(4\*a+4+a^2\*\rho)}
   \Big[8+16\*a-a^4\*\rho-3\*a^3\*\rho^2+10\*a^2\*\rho+6\*a^3\*\rho+4\*\rho\* a 
   + 8\*a^2 \Big]\*L_1\nonumber\\[2mm]
   & +\frac{2}{(a+1)\*(4\*a+4+a^2\*\rho)}\Big[8\*a^2+16\*a +8+4\*a^2\*\rho
            -a^3\*\rho^2 + 4\*a^3\rho\Big]\*L_2\nonumber\\[2mm]
      &-\frac{8\*\beta}{(4\*a+4+a^2\*\rho)} 
      \Big[a^2\*\rho+2\*a+2\Big]\*L_3
      +\frac{2 \*\beta}{(4\*a+4+a^2\*\rho)}\*\Big[a^2\*\rho+10\*a+10\Big]
      - 4 \* L_{10} + a \* \rho \* L_2^2
      \biggr\},
\end{align}
\begin{align}
L_1 &= \log\left(\tfrac{(1-\beta)\*(a\*(1+\beta)+2)}{(1+\beta)\*(a\*(1-\beta)+2)}\right),&\quad
L_2 &= \log\left(\tfrac{1+\beta}{1-\beta}\right), \quad\quad
L_3  = \log\left(\tfrac{4\beta^2}{\rho}\right),\nonumber\\[2mm]
L_4 &= \Li\left(\tfrac{1-\beta}{2}\right) - \Li\left(\tfrac{1+\beta}{2}\right),&\quad
L_5 &= \Li\left(-\tfrac{2\*\beta}{1-\beta}\right) -
         \Li\left(\tfrac{2\*\beta}{1+\beta}\right),\nonumber\\[2mm]
L_6 &=  \log^2\left(1 + \beta\right) - \log^2\left(1 - \beta\right),&\quad
L_7 &=  \Li\left(\tfrac{a\*(1-\beta)}{2\*(1+a)}\right) -
       \Li\left(\tfrac{a\*(1+\beta)}{2\*(1+a)}\right),\nonumber\\[2mm]
L_8 &= \Li\left(\tfrac{-a\*(1+\beta)}{2}\right) -
         \Li\left(-\tfrac{a\*(1-\beta)}{2}\right),&\quad
L_9 &= \Li\left(-\tfrac{2 \*a \*\beta}{a\*(1-\beta)+2}\right) - 
       \Li\left(\tfrac{2 \*a \*\beta}{a\*(1+\beta)+2}\right),\nonumber\\[2mm]
L_{10} &=\Li\left(-\tfrac{2\*\beta}{1-\beta}\right) +
         \Li\left(\tfrac{2\*\beta}{1+\beta}\right),&\quad
\Li(x) &= -\int_{0}^{x} \mathrm{d}t \frac{\log(1 - t)}{t},\quad\quad
 a = \frac{\mg^2}{\mq^2} - 1\quad .
\end{align}
The leading order scaling functions $f_{gg}^{(00)}$ and $f_{q\barq}^{(00)}$ can be found in
Ref.~\citep{Langenfeld:2009eg}.
The NNLO scale dependent scaling functions follow from the RGE relations
\begin{align}
 \label{eq:f21}
f_{ij}^{(21)} \,=\, & 
\frac{1}{(16\pi^2)^2}\*\left(
  2\* \beta_1\* f_{ij}^{(00)} - f_{kj}^{(00)}\otimes P_{ki}^{(1)} 
  - f_{ik}^{(00)}\otimes P_{kj}^{(1)}\right)
+ \frac{1}{16\pi^2}\*\left(
  3 \*\beta_0 \*f_{ij}^{(10)} 
  -f_{kj}^{(10)}\otimes P_{ki}^{(0)}
  - f_{ik}^{(10)}\otimes
  P_{kj}^{(0)}\right)
\, ,
\\[2mm]
\label{eq:f22}
f_{ij}^{(22)} \,=\,&
\frac{1}{(16\pi^2)^2}\*\left(
  f_{kl}^{(00)}\otimes P_{ki}^{(0)}\otimes P_{lj}^{(0)}
  +\frac{1}{2} f_{in}^{(00)}\otimes P_{nl}^{(0)}\otimes P_{lj}^{(0)}
  +\frac{1}{2} f_{nj}^{(00)}\otimes P_{nk}^{(0)}\otimes P_{ki}^{(0)} 
  + 3 \*\beta_0^2  \*f_{ij}^{(00)}\right.\notag\\*[2mm]
&  \hspace{18mm}\left.
  - \frac{5}{2}\*\beta_0 \*f_{ik}^{(00)}\otimes P_{kj}^{(0)}
  - \frac{5}{2}\*\beta_0 \*f_{kj}^{(00)}\otimes P_{ki}^{(0)}
\right)
\end{align}
$i,j,k,l,n$ are parton indices with implied summation over repeated indices.
The threshold expansion is derived by computing the Mellin transformation of each of the
involved factors and inverting the products back to $\rho$ space.
The scaling function $f^{(ij)}_{gq}$ is very small near threshold 
so we did not include them.
The constants $a_1^{qq,gg}$ can be found in Ref.~\citep{Langenfeld:2009eg}.
The coefficients of the QCD $\beta$ - function are given as $\beta_0 = 11 -(2/3)\nf$ 
and $\beta_1 = 102 -(38/3)\nf$. 
We used the threshold expansion to fit the numerical determined values of the NNLO
scaling functions.
\begin{align}
f^{(21)}_{gg} = & \frac{f^{(00)}_{gg}}{(16\pi^2)^2}\Big[-4608\log^3(\beta)
                   +\Big(-18432\*\log(2)+
                    \tfrac{109920}{7}-64\*\nf\Big)\log^2(\beta)\nonumber\\[0mm]
                 & \hspace*{14mm}+\Big(6766.94811-66178.09806\*\agg
                       -192\log(2)\*\nf+\tfrac{4048}{21}\*\nf
                        -\tfrac{176}{7}\*\tfrac{\pi^2}{\beta}\Big)\log(\beta)\nonumber\\[1mm]
                  & \hspace*{14mm} -3572.87371+35472.75010 \agg +
                                 55.41606408\*\nf-919.1402509\*\agg\*\nf\nonumber\\[1mm]
                  &  \hspace*{14mm}+\tfrac{56.86772061}{\beta}-\tfrac{3.446528522\*\nf}{\beta}
                 \Big]\\[3mm]
f^{(22)}_{gg} = & \frac{f^{(00)}_{gg}}{(16\pi^2)^2}\Big[
               1152\*\log^2(\beta)+\Big(16\*\nf-2568+2304\*\log(2)\Big)\*\log(\beta)
               \nonumber\\[0mm]
           &\hspace*{14mm} 2568+1152 \log^2(2)-2568 \log(2)-144\pi^2 
           + 16 \nf \log(2)-16 \nf\Big]\\[3mm]
f^{(21)}_{q_{i}\bar{q}_{j}} = &
     \frac{f^{(00)}_{q_{i}\bar{q}_{j}}}{(16\pi^2)^2}\Big[-\tfrac{8192}{9}\log^3(\beta)
                        +\Big(-\tfrac{32768}{9}\*\log(2)+\tfrac{34688}{9}
                        -\tfrac{256}{3}\*\nf\Big)\log^2(\beta)\nonumber\\[0mm]
               &\hspace*{14mm} +\Big(1150.2835-2412.743158\*\aqq +\tfrac{5392}{27} \nf 
              -256\*\log(2)\*\nf-\tfrac{448}{9}\*\tfrac{\pi^2}{\beta}\Big)\log(\beta)
              \nonumber\\[1mm]
               &\hspace*{14mm}-1374.416616+3567.790429 \aqq +70.72319322 \nf
                            -226.1946711\*\aqq\*\nf\nonumber\\[1mm]
               &\hspace*{14mm} + \tfrac{9235}{\beta}- \tfrac{46.05815389\*\nf}{\beta}
               \Big]\\[3mm]
f^{(22)}_{q_{i}\bar{q}_{j}} = & \frac{f^{(00)}_{q_{i}\bar{q}_{j}}}{(16\pi^2)^2}\Big[
                        \tfrac{2048}{9}\log^2(\beta)
                        +\Big(-\tfrac{7840}{9}+\tfrac{320}{9} \nf 
                        +\tfrac{4096}{9} \log(2)\Big)\log(\beta)\nonumber\\[0mm]
                      &\hspace*{14mm} +\tfrac{9415}{9} + \tfrac{2048}{9} \log^2(2) 
                        -\tfrac{7840}{9} \log(2) - \tfrac{256}{9} \pi^2
                        -\tfrac{596}{9}\*\nf+\tfrac{320}{9} \log(2)\nf
                        +\tfrac{4}{3} \nf^2
                        \Big]
\end{align}

\begin{theacknowledgments}
I would like to thank S. Moch for reading the manuskript and P. Falgari for pointing to
some errors in Eq.~(\ref{eq:fgg11}) and (\ref{eq:fqq11}).
This work is supported by the Helmholtz Gemeinschaft under contract VH-NG-105
and by the Deutsche Forschungsgemeinschaft under contract SFB/TR 9.
\end{theacknowledgments}


\end{document}